\documentclass[useAMS,usenatbib]{mn2e}

\usepackage{graphicx}

\title[A decelerating jet in the X-ray transient XTE~J1752$-$223]{A decelerating jet observed by the EVN and VLBA \\ in the X-ray transient XTE~J1752$-$223}
\author[J. Yang, C. et al.]{
J.\,Yang$^{1}\thanks{E-mail: yang@jive.nl}$,
C.\,Brocksopp$^{2}$,
S.\,Corbel$^{3}$,
Z.\,Paragi$^{1,4}$,
T.\,Tzioumis$^{5}$
and R.P. Fender$^{6}$
\\
$^{1}$Joint Institute for VLBI in Europe, Postbus 2,  7990 AA Dwingeloo, The Netherlands \\
$^{2}$Mullard Space Science Laboratory, University College London, Holmbury St Mary, Dorking, Surrey RH5~6NT, UK \\
$^{3}$Universit\'e Paris 7 Denis Diderot and Service d'Astrophysique, UMR AIM, CEA Saclay, F-91191 Gif-sur-Yvette, France \\
$^{4}$MTA Research Group for Physical Geodesy and Geodynamics, POB 91, H-1521 Budapest, Hungary \\
$^{5}$Australia Telescope National Facility, CSIRO, P.O. Box 76, Epping, NSW 1710, Australia. \\
$^{6}$School of Physics and Astronomy, University of Southampton, Highfield, Southampton, SO17~1BJ, UK }

\begin{document}

\date{Accepted 2010 September 7. Received 2010 August 27; in original form 2010 July 23}

\pagerange{\pageref{firstpage}--\pageref{lastpage}} \pubyear{2010}

\maketitle

\label{firstpage}

\begin{abstract}
The recently discovered Galactic X-ray transient XTE~J1752$-$223
entered its first known outburst in 2010, emitting from the X-ray to the radio
regimes. Its general X-ray properties were consistent
with those of a black hole candidate in various spectral states, when
ejection of jet components is expected. To verify this, we
carried out very long baseline interferometry (VLBI) observations.
The measurements were carried out with the European VLBI Network
(EVN) and the Very Long Baseline Array (VLBA) at four epochs in 2010 February.
The images at the first three epochs show a moving jet
component that is significantly decelerated by the last epoch,
when a new jet component appears that is likely to be associated with
the receding jet side. The overall picture is consistent with
an initially mildly relativistic jet, interacting with the
interstellar medium or with swept-up material along the jet.
The brightening of the receding ejecta at the final epoch can be
well explained by initial Doppler deboosting of the emission
in the decelerating jet.

\end{abstract}


\section{Introduction}
It is clear from the literature of the past three decades that, for almost every black hole X-ray transient (XRT) observed at radio wavelengths, a radio counterpart has been discovered \citep{fen06}. In a small number of sources, the ejecta have been resolved and monitored as they travel away from the central source. Thus, it has been possible on rare occasions to measure proper motions, sometimes at apparently super-luminal velocities: e.g. GRS~1915$+$105 \citep{mir94, fen99}, GRO~J1655$-$40 \citep{tin95, hje95}, GX~339$-$4 \citep{cor10}, and even detect the only known instances of a Galactic parsec-scale jet being decelerated in the ISM: XTE~J1550$-$564 \citep{cor02, kaa03} and XTE~J1748$-$288 \citep{hje98, mil08}. Jet deceleration has been also investigated in GRS~1915$+$105 although no conclusive evidence was found \citep{mil07}.

The X-ray transient XTE~J1752$-$223 was discovered by the \emph{Rossi X-ray Timing Explorer} (\emph{RXTE}) on 2009 October 23 \citep{mar09} at the start of its first known X-ray outburst. It showed a long and gradual rise at X-ray energies, whilst remaining spectrally hard. The X-ray source later evolved, became softer \citep{hom10,bro10a} and entered a spectral state commonly associated with jet ejection events \citep{fen04, fen09}. The outburst has been well monitored by the \emph{Monitor of All-sky X-ray Image} \citep[\emph{MAXI},][]{nak10}, \emph{RXTE} \citep{sha10}, and \emph{Swift} \citep{cur10} at X-ray energies. All these X-ray observations show that XTE J1752$-$223 is likely to be a black hole transient.

Following the activation of XTE~J1752$-$223, we initiated the ATCA (Australia Telescope Compact Array) radio observations and discovered the radio counterpart with a flux density of $\sim 2$~mJy at 5.5~GHz \citep{bro09}. The later ATCA monitoring observations showed that the radio source entered a series of flares, peaking at 5~--~20 mJy (Brocksopp et al. in prep.), after the transition from X-ray hard to soft state. To detect the potential ejecta and study their evolution, we carried out an EVN (European VLBI Network) experiment and three follow-up VLBA experiments at 5 GHz in 2010 February. In this paper, we present the results of these VLBI (Very Long Baseline Interferometry) observations.

\section{Rapid EVN and VLBA Observations}
The performed VLBI experiments are summarised in Table~\ref{tab1}. The low declination and potentially weak flux of XTE~J1752$-$223 were possible problems for VLBI observations. Therefore, to quickly resolve these concerns, we initiated an e-VLBI experiment with the western European telescopes \citep{szo08}. The EVN experiment used 1024~Mbps data rate (16 channels, 16~MHz per channel, 2~bit sampling, dual polarisation). The real-time correlation was done with the Earth orientation parameters (EOP) predicted from the EOP model of one day earlier. We applied 2-second integration time, and 16 frequency points per channel. The participating stations were Medicina, Yebes, Torun, Onsala, and Westerbork. In the EVN experiment, we used the following J2000 coordinate: $\mathrm{RA}=17^\mathrm{h}52^\mathrm{m}15\fs095$, $\mathrm{DEC}~=-22\degr20\arcmin32\farcs782$ (positional uncertainty: $\sigma=0\farcs3$), which was determined by the ATCA observations \citep{bro09}. To obtain fringe-fitting solutions and a reference point, we scheduled a nearby ($0\fdg8$) phase-referencing source: PMN~J1755$-$2232
($\mathrm{RA}=17^\mathrm{h}55^\mathrm{m}26\fs285$, $\mathrm{DEC}=-22\degr32\arcmin10\farcs593$, J2000, position error $\sigma\sim15$~mas).  As these sources have low elevation ($<30\degr$) in Europe, we used a short cycle time: 160 seconds on the target and 80~seconds on the reference source. We also observed a strong and compact source (NRAO~530) as the fringe finder and bandpass calibrator.

We successfully detected a radio source, consistent with an ejection from the black hole candidate \citep{bro10b}, during the EVN experiment and then performed three follow-up VLBA experiments. We used the same phase-referencing source, cycle time, and observing frequency. The recording data rate was 512~Mbps (16~channels, 8~MHz per channel, 2~bit sampling, dual polarisation). There were eight VLBA telescopes available at the first epoch, nine at other two epochs. The data were correlated with the same parameters as the EVN experiment.

We also performed an EVN experiment in March and two VLBA experiments in April to study other ejection events and attempt to detect the core. These additional results will be presented by Brocksopp et al. in a general paper.

\section{VLBI Data Calibration}
We used the NRAO software {\sc AIPS} (Astronomical Image Processing System) to perform the initial calibrations. The \emph{a-priori} amplitude calibration was done with measured system temperatures and antenna gain curves. We corrected the EOP model for the VLBA data before any phase calibrations. We follow the same procedure for both the EVN and VLBA data reduction. (1) We corrected the parallactic angle. (2) We ran the global fringe-fitting for NRAO~530 with half-minute solution interval and then solved for the instrumental bandpass. (3) With the bandpass solutions, we re-ran the fringe-fitting to solve for the instrumental phase and delay using two-minute data of NRAO~530. (4) We ran the fringe-fitting to solve for the phase, the fringe rate, and the delay for the calibrators with a solution interval of the scan length ($\sim1$ minute). On the long baselines to Mauna Kea (Mk) and St. Croix (Sc), there were no solutions found for PMN~J1755$-$2232 as it has much lower correlation amplitude ($<10$~mJy) most likely due to scatter broadening. In view of this problem, we removed both Sc and Mk. The phase wrapped slowly (fringe rate $<5$~milliHz). The solutions of PMN~J1755$-$2232 were then transferred to XTE~J1752$-$223 by linear interpolation. (5) The data were averaged in each IF and then split into single-source files after all the corrections were applied.

The reference source PMN~J1755$-$2232 was imaged by circular Gaussian model fitting and self-calibration in {\sc Difmap}\citep{she94}. The source was well represented by a circular Gaussian model with a size of $4.2$~mas and a total flux density of $\sim200$~mJy at 5~GHz. Finally, we self-calibrated the $u$--$v$ data with the model and applied the amplitude and phase solutions to both sources in AIPS.

\begin{figure*}
  \centering
  \includegraphics[width=0.71\textwidth]{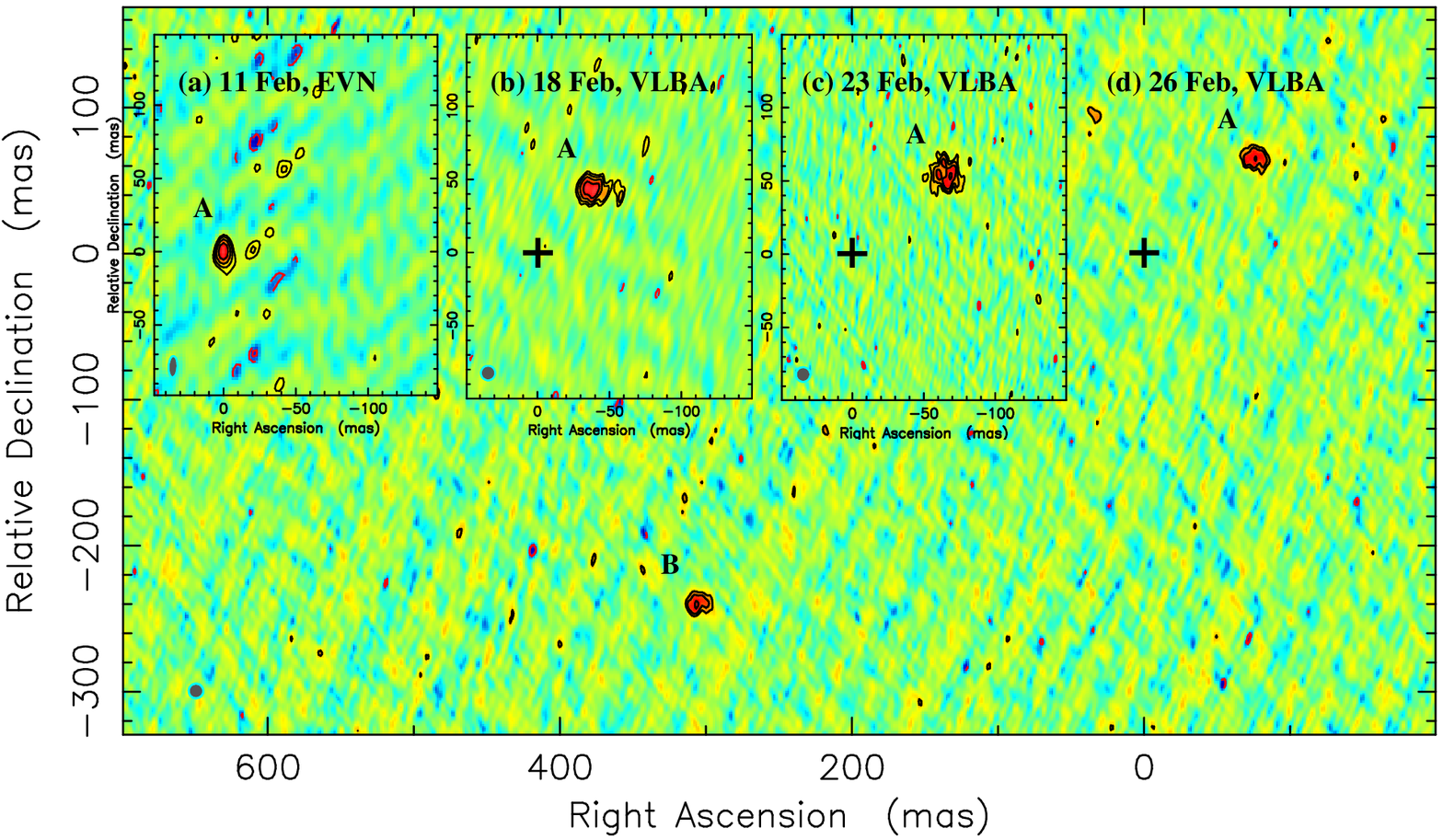}
  \includegraphics[width=0.27\textwidth]{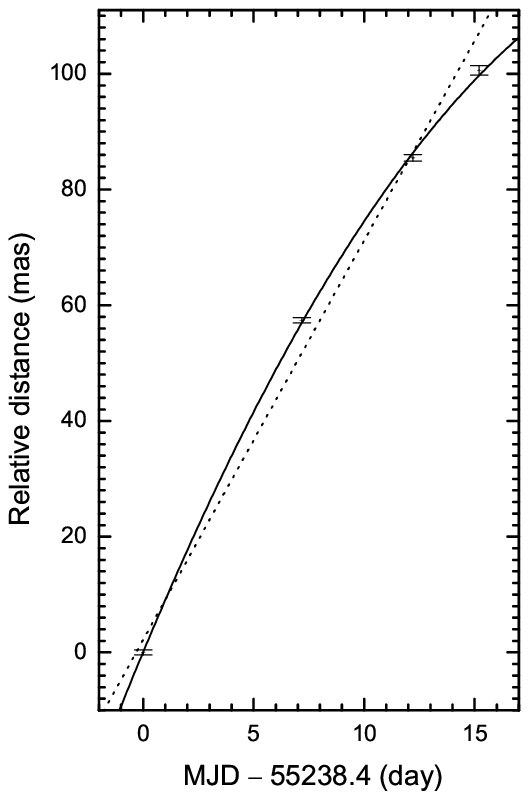} \\

  \caption{The decelerating jet of the X-ray transient XTE~J1752$-$233. All the VLBI images are centred at the location of component A on 2010 February 11, indicated by a cross. Component B was not detected until 2010 February 26. The contours start from $3\sigma$ off-source noise level and increase by a factor of -1.4, -1, 1, 1.4, 2, 2.8. The related map parameters are listed in Tab.~\ref{tab1}. The right panel plots the fitting results (Tab.~\ref{tab3}) of the proper motion data (Tab.~\ref{tab2}) of component A using the models with (solid curve) and without deceleration rate (dotted line). The reference time MJD 55238.4 corresponds to the first VLBI observations.} \label{fig1}
\end{figure*}

\begin{table}
  \caption{\small The summary of the image parameters of Fig.~\ref{fig1}.
}\label{tab1}
\tiny
\setlength{\tabcolsep}{1.5pt}
\begin{tabular}{lccccclrrr}
\hline
Exp.   & Date     & Array  & $N_\mathrm{ant}$
                                   & $T_\mathrm{obs}$
                                             & $S_\mathrm{peak}$
                                                      & $\sigma_\mathrm{rms}$
                                                                & $b_\mathrm{maj}$
                                                                         & $b_\mathrm{min}$
                                                                                  & $\phi_\mathrm{pa}$  \\
       & dd/mm/yy   &       &      & (hour)  & (mJy/b)& (mJy/b)  &  (mas) & (mas) & ($\degr$)\\
\hline
RY001  & 11/02/10   & EVN   & 5    &  1.2    & 2.32   &   0.21   & 14.5   & 6.1   & $-3$     \\
BB290A & 18/02/10   & VLBA  & 6    &  3.0    & 0.77   &   0.072  & 10.0   & 10.0  &  0       \\
BB290B & 23/02/10   & VLBA  & 7    &  6.0    & 0.60   &   0.068  & 10.0   & 10.0  &  0       \\
BB290C & 26/02/10   & VLBA  & 7    &  6.0    & 0.37   &   0.057  & 10.0   & 10.0  &  0       \\
\hline
\end{tabular}

\medskip
The columns give: (1) experiment code; (2) date; (3) array name; (4) total number of the used telescopes; (5) total observing time; (6) peak flux density; (7) off-source noise level; (8 -- 10) sizes of the major and minor axes of restoring beam and its position angle.\\
\end{table}

\section{VLBI detection of XTE~J1752$-$223}

The imaging results for the X-ray transient XTE~J1752$-$223 are shown in the left panel of Fig.~\ref{fig1}. The top small panels from left to right are the EVN image of 2010 February 11 and the VLBA images of 2010 February 18 and 26. The background large image is the VLBA image of 2010 February 26. The cross in each image marks the position of component A at the first epoch ($\mathrm{RA}=17^\mathrm{h}52^\mathrm{m}15\fs06370$, $\mathrm{DEC}=-22\degr20\arcmin31\farcs9838$, J2000). At the bottom of the large background image, there is another jet component marked as component B at an angular separation $488$~mas from component A and at a position angle consistent with the moving direction of component A. The related map parameters are listed in Tab.~\ref{tab1}.

Component A, surrounded with the beam pattern, is clearly seen with a peak brightness of 2.32~mJy~beam$^{-1}$ in the dirty map at the first epoch, when natural weighting is used. After removing component A with a circular Gaussian model, we notice that there may be at least one more jet component. One candidate is located at angular separation 18.7~mas, position angle $-84\fdg0$; the other at angular separation 70.6~mas and position angle $-36\fdg3$. Both candidates have a peak brightness $\sim0.91$~mJy~beam$^{-1}$ ($\sim4\sigma_\mathrm{rms}$) using natural weighting. In Fig.~\ref{fig1}a, the two candidates show the second positive contours. If either candidate is removed by circular Gaussian model fitting, the other also becomes faint. If we reduce the contribution of the long baselines, both become brighter and show a small peak ($\sim5$\%) brightness difference. Due to the limited sensitivity and $u$~--~$v$ coverage during the 1.2-hour observations, neither components can be unambiguously identified as a true jet component. However, there is evidence for the extended emission for the source as the total restored flux density is much lower than that ($\sim16$~mJy at 5.5~GHz) measured by the ATCA (Brocksopp et al. in prep.).

In the follow-up VLBA observations, the higher resolution and sensitivity are achieved by more telescopes and longer observing time. To image the extended source, we used natural weighting again. Because of the resolved structure and the decaying peak flux density, the source is only seen clearly in the dirty map with the synthesised beam ($16.2\times3.9$~mas at position angle $-15\fdg6$) at the first VLBA epoch. However, the large-scale beam pattern around the faint source could also be easily identified at the later two epochs. If we taper the long baselines, use the short baselines only, or increase the image pixel size, the source becomes significantly brighter in the dirty map at the later two epochs. None of the suspected ejecta candidates in the EVN image are further seen after 7 days in the later VLBA images. Because the diffuse emission can not be well restored by clean components, Gaussian models were used in making all the VLBI images of Fig.~\ref{fig1}. Due to the limited SNR (6~--~12), circular rather than elliptical Gaussian model fitting was adopted to reduce the number of free parameters.

Table~\ref{tab2} lists the best-fit parameters of the circular Gaussian model. To show the motion of component A, we take the position of component A measured at the first epoch as the reference origin. The random position error was estimated by $\frac{\sqrt{b_\mathrm{maj}b_\mathrm{min}}}{2\mathrm{SNR}}$, where $b_\mathrm{maj}$ and $b_\mathrm{min}$ are the size of the major and minor axes of the used restoring beam, $\mathrm{SNR}$ is the signal to noise ratio ($\frac{S_\mathrm{peak}}{\sigma_\mathrm{rms}}$) listed in Column (7) of Table~\ref{tab2}. Note that the rather large systematic position error from the reference source will not affect our proper motion measurements. The fitted size has the same random error as the position for each component. Since the measured sizes ($\geq8$~mas) are much larger than that (4.2~mas) of the reference source, they should be very close to the true size of the ejecta. At the second epoch, we notice that component A shows an elongated structure in the East-West direction and the eastern side is brighter than the western side. This brightness distribution caused a slightly different position angle of the component, compared to what is measured at later epochs. The VLBI flux density errors are usually  $\sim5\%$.

\begin{table}
\caption{\small The circular Gaussian model fitting results of the detected jet components in the X-ray transient XTE~J1752$-$223.} \label{tab2}
\tiny
\setlength{\tabcolsep}{5pt}
\centering
\begin{tabular}{lcrrrcr}
\hline
Comp. & MJD      & Separation       & Position Angle      & Size    & Flux      & SNR    \\
      & (day)    & (mas)            &($^\circ$)           & (mas)   & (mJy)     &        \\
\hline
A     & 55238.4  & $0\pm0.4$        &  0                  &   7.9   & 4.35      &  11.0  \\
A     & 55245.6  & $57.4\pm0.5$     &  $-41.18\pm0.90$    &  13.8   & 2.20      &  10.7  \\
A     & 55250.6  & $85.5\pm0.6$     &  $-50.14\pm0.63$    &  19.0   & 2.32      &   8.8  \\
A     & 55253.6  & $100.6\pm0.8$    &  $-49.23\pm0.70$    &  13.9   & 1.05      &   6.1  \\
B     & 55253.6  & $387.9\pm0.8$    &  $128.08\pm0.18$    &  11.9   & 0.86      &   6.4  \\
\hline
\end{tabular}
\\
\end{table}

\section{Gradual jet deceleration}
The angular separation of component A versus time is shown in the right panel of Fig.~\ref{fig1}. We take the position and the time of component A measured at the first epoch as the reference origin. We fit these data points to the following proper motion model:
\begin{equation}\label{eq1}
r=r_0+\mu_0 t - 0.5\dot\mu t^2
\end{equation}
where $r$ is the angular separation; $t$ is the observing time; $r_0$ and $\mu_0$ are the angular separation and the proper motion at $t=0$, $\dot\mu$ is the apparent deceleration rate. The dotted straight line and the solid curve represent the uniform proper motion model ($\dot\mu=0$) and the proper motion model with the deceleration rate ($\dot\mu\neq0$) respectively. The best-fit parameters are listed in Table~\ref{tab3}. The model of $\dot\mu=0$ gives an average proper motion of $\bar\mu=\mu_0=6.90\pm0.05$~mas~day$^{-1}$ with the reduced $\chi^2=118.4$. The
degree of freedom (DoF) was listed in the last column. The model of $\dot\mu\neq0$ gives $\mu_0=9.15\pm0.15$~mas~day$^{-1}$ at MJD~55238.4 and a deceleration rate of $\dot\mu=0.34\pm0.02$~mas~day$^{-2}$ with the reduced $\chi^2=3.9$. With the deceleration rate, component A has a proper motion of 4.0~mas~day$^{-1}$ at the last epoch. It is clear that the deceleration rate should be included in the proper motion model.

Jet deceleration was also found in XTE~J1550$-$564 using \emph{Chandra} observations \citep{cor02, kaa03} and XTE~J1748$-$288 with VLA observations \citep{hje98, mil08}. Compared with them, the observed deceleration in XTE~J1752$-$223 is free from the blending of multiple jet components caused by the low resolution \citep{hje95}. If there is a sequence of ejecta which decreased sequentially more rapidly in flux density with increasing distance from the core, the cluster of components may show a decreasing proper motion. In our case, these VLBI observations have a resolution of $<10$~mas and can well identify single ballistic ejecta with a time resolution of less than one day, assuming an initial proper motion 10~mas~day$^{-1}$. XTE~J1752$-$223 is the second known case of gradual jet deceleration, although on a much smaller scale ($\sim100$~mas) than the first case of XTE~J1550$-$564. As for the previous two sources, the jet deceleration in XTE~J1752$-$223 is most likely due to interaction with the external dense interstellar medium (ISM) or the residual slowly-moving ejecta from the previous ejection along the jet path.

\begin{table}
\caption{\small Best-fit parameters using the proper motion models with and without the deceleration rate. }\label{tab3}
\scriptsize
\setlength{\tabcolsep}{4.5pt}
\centering
\begin{tabular}{llllrl}
\hline
Model            & $r_{_0}$       & $\mu_{_0}$          & $\dot\mu$        &$\chi^2$/DoF & DoF     \\
                 & (mas)          & (mas~day$^{-1})$    & (mas~day$^{-2}$) &             &         \\
\hline
$\dot\mu\neq0$   & $0.06\pm0.41 $ & $9.15\pm0.15$       & $0.34\pm0.02$    & 3.9         &  1      \\
$\dot\mu=0$      & $2.16\pm0.39 $ & $6.90\pm0.05$       &  0               & 118.4       &  2      \\
\hline
\end{tabular}
\end{table}

\section{Component B: the receding ejecta?}
\begin{figure}
  \centering
  \includegraphics[height=3.2cm]{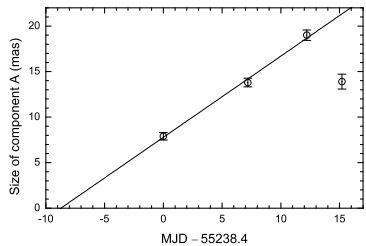}\\
  \caption{Evolution of the size of component A. The solid straight line shows the fitting result of the first three data points.
  }\label{fig2}
\end{figure}

We interpret component A as an approaching jet component since it is the only component detected at the four epochs and our VLBI observations were performed just after the associated radio flare reached its peak flux density. The ATCA observations (Brocksopp et al., in prep.) show that it had a decaying flux density and a pretty stable and steep spectral index: $\alpha=-1$ ($S_\nu\propto\nu^{\alpha}$) between 5.5 and 9~GHz, i.e. there was no indication of another ejection event during our VLBI observations. Thus, the possibility that the components A and B detected at the last three epochs are associated with a different ejection event can be ruled out. Besides the proper motion, component A shows a hint of expansion. The evolution of its size is displayed in Fig.~\ref{fig2}. The first three data points give an expansion speed of $0.9\pm0.1$~mas~day$^{-1}$ with reduced $\chi^2=1.1$. The expansion speed is much slower than its proper motion, indicating that its expansion was significantly confined too. Because component A shows an increasing size and a decaying peak brightness, its size estimation at a later stage is limited by the image sensitivity and the lack of short baselines. For this reason, the last data point was omitted in the linear fitting. According to the evolution of component A, a jet component is expected to have more compact structure at the earlier stage. Compared with the size (7.9~mas) of component A measured at the first epoch, component B shows a much larger size (11.9~mas). This indicates that component B is most likely an evolved component, which was ejected on the receding jet side at the same time as component A.

According to the expansion speed, component A was ejected 8.7~days earlier, i.e. at MJD~$55229.7$ (2010 Feb. 2), which is represented by the x-intercept in Fig.~\ref{fig2}. As component A may have significantly larger expansion speed and strong Doppler boosting effect if it is unhindered at this earlier stage, the inferred birth date may be the earliest possible birth date. Although such extrapolation is not guaranteed, the inferred date is at the beginning of the initial rising stage of the associated radio flare in the ATCA radio light curve (Brocksopp et al. in prep.). The average separation speed of the pair of components is $\bar\mu_\mathrm{app}+\bar\mu_\mathrm{rec}\geq20.4$~mas~day$^{-1}$ if they were indeed ejected on the inferred birth date. Since $\bar\mu_\mathrm{app}\geq\bar\mu_\mathrm{rec}$, there is $\bar\mu_\mathrm{app}\geq10.2$~mas~day$^{-1}$, which is significantly higher than the average proper motion (6.9~mas~day$^{-1}$) measured during our observations. Thus, component A had already been significantly decelerated before our VLBI observations.

If the jet expansion is linear and symmetric on both sides, the ratio of the approaching and receding component sizes is \citep[e.g.][]{mil04}:
\begin{equation}\label{eq2}
\frac{R_\mathrm{app}(t_\mathrm{app})}{R_\mathrm{rec}(t_\mathrm{rec})}=\frac{t_\mathrm{app}}{t_\mathrm{rec}}=\frac{1+\bar\beta(0,t_\mathrm{app})\cos\theta}{1-\bar\beta(0,t_\mathrm{rec})\cos\theta}
\end{equation}
where $t_\mathrm{app}$ and $t_\mathrm{rec}$ are the intrinsic times at which light leaves the approaching and receding jet components respectively and arrive at the telescope at the same observing time, $\bar\beta$ is the average jet speed in units of light speed $c$ and $\theta$ is the inclination angle of the jet axis. By a linear extrapolation, the approaching jet component has a size of 21.3~mas at the fourth epoch. If there is no deceleration, $\bar\beta(0,t_\mathrm{app}) = \bar\beta(0,t_\mathrm{rec})=\beta$, we can give $\beta\cos\theta=0.3c$, which requires $\beta\geq0.3c$ and $\theta\leq73\degr$. Note that the size of the receding jet component detected for the first time is not likely to be affected by the over-resolution since it is much younger than the approaching jet component ($t_\mathrm{rec}\sim0.5t_\mathrm{app}$). Given the jet deceleration and $t_\mathrm{rec}\leq t_\mathrm{app}$ at the same telescope time, then $\bar\beta(0,t_\mathrm{rec})\geq0.3c$ and $\bar\beta(0,t_\mathrm{app})\leq0.3c$. Therefore, we can take $0.3c$ as the lower limit of the jet birth speed in the case of the jet deceleration.

The ratio of the flux density measured at $R_\mathrm{app}=R_\mathrm{rec}$ ($t_\mathrm{app}=t_\mathrm{rec}$, free from its intrinsic luminosity evolution effect) for a pair of discrete jet components \citep{mir99}:
\begin{equation}\label{eq3}
    \frac{S_\mathrm{app}}{S_\mathrm{rec}} = \left( \frac{1+\beta\cos\theta}{1-\beta\cos\theta} \right)^{3-\alpha}
\end{equation}
The receding jet component had a size of $R_\mathrm{rec}=11.9$~mas at the last epoch. The corresponding observing time for the approaching jet component is at MJD~55242.9 (between the first two epochs).  At $t_\mathrm{app}=t_\mathrm{rec}$, the radio core is inferred to be at the centre: angular separation $\sim174$~mas and position angle $\sim-130\degr$. The radio core was not detected during any of the VLBI epochs but, since all four observations took place during the X-ray soft state, this is to be expected, according to the unified model of \citet{fen04, fen09}. The radio position confirms that its optical counterpart is \emph{Swift}-UVOT source A \citep{cur10}. If we take the flux density at the first epoch as the upper limit of $S_\mathrm{app}$, then $\frac{S_\mathrm{app}}{S_\mathrm{rec}}\leq5$ and $\beta\cos\theta\leq0.2c < \bar\beta(0,t_\mathrm{rec})\cos\theta$, in agreement with the jet deceleration scenario on both sides. The observed flux density from the receding jet is deboosted by a factor: $\delta_\mathrm{rec}^{3-\alpha}$, where $\delta_\mathrm{rec}=(1-\beta^2)^{0.5}(1+\beta\cos\theta)^{-1}$. Because of the jet deceleration, the receding ejecta is less beamed away from our line of sight, and thus it looks relatively brighter. Note that the non-detection of the receding jet component at the first epoch may also be because it stayed at an earlier stage ($t_\mathrm{rec}\sim0.5t_\mathrm{app}$ if $\beta\cos\theta=0.3$) and its flux density was still much lower than the peak flux density of the radio flare. The caveat in the above argument is that component B might have not followed the same luminosity evolution model as component A. It is also possible that the brightening of the receding ejecta was due to sudden jet-cloud interaction, as in the case of XTE~J1748$-$228 \citep{hje98, mil08}.

\section{Conclusions}
In this paper, we present the results of the first VLBI observations of the new Galactic black hole candidate XTE~J1752$-$223 during its first known outburst. With EVN and VLBA observations at four epochs in 2010 February, we imaged its radio counterpart at 5~GHz. We detect an ejected component at the first three epochs and find that its proper motion shows significant deviation from the uniform proper motion model and requires a deceleration rate of $0.34\pm0.02$~mas~day$^{-2}$. In the jet deceleration scenario, it has proper motion decelerating from 9.2~mas~day$^{-1}$ at the first epoch to 4.0~mas~day$^{-1}$ at the last epoch. It also shows slow but detectable variation of its transverse size indicating that its expansion is also significantly confined. This is the first time that a Galactic jet is found to be gradually decelerating on the hundred milliarcsecond scale. The discovery provides strong evidence for the existence of significant interaction around the jet at an early stage of its evolution. In addition to the approaching jet component, we detect another jet feature at the last epoch, which is most likely associated with the receding ejecta. We infer that the jet deceleration should start at a time much earlier than our VLBI observations using the birth date (around 2010 February 2) from the ATCA radio light curve. Furthermore, we interpret the detection of the receding ejecta as a result of the reduced Doppler deboosting effect caused by the jet deceleration on the receding side and give a lower limit of $0.3c$ for the jet birth speed assuming symmetric jet motion. It has been reported by \citet{sha10} that the distance, estimated by the spectral-timing correlation scaling technique, is around 3.5~kpc. Thus, the proper motion observed at the first epoch would correspond to an apparent jet speed of $\sim0.2c$, in agreement with our results (but note that the technique is very model dependent).

\section*{Acknowledgments}
\footnotesize
We thank the EVN PC Chair, T. Venturi and the VLBA Proposal Selection Committee for prompt approval of our ToO requests. e-VLBI developments in Europe were supported by the EC DG-INFSO funded Communication Network Developments project EXPReS. The National Radio Astronomy Observatory is a facility of the National Science Foundation operated under cooperative agreement by Associated Universities, Inc. The European VLBI Network is a joint facility of European, Chinese, South African and other radio astronomy institutes funded by their national research councils.


\label{lastpage}


\normalsize

\begin{thebibliography}{99}

\bibitem[\protect\citeauthoryear{Brocksopp et al.}{2009}]{bro09} Brocksopp~C. et al., 2009, Astronomer's Telegram,~2278

\bibitem[\protect\citeauthoryear{Brocksopp et al.}{2010a}]{bro10a} Brocksopp~C. et al.,  2010, Astronomer's Telegram,~2400

\bibitem[\protect\citeauthoryear{Brocksopp et al.}{2010b}]{bro10b} Brocksopp~C. et al., 2010, Astronomer's Telegram,~2438

\bibitem[\protect\citeauthoryear{Corbel et al.}{2002}]{cor02} Corbel~S. et al., 2002, Sci, 298, 196

\bibitem[\protect\citeauthoryear{Corbel et al.}{2010}]{cor10} Corbel~S. et al., 2010, Astronomer's Telegram, 2745

\bibitem[\protect\citeauthoryear{Curran et al.}{2010}]{cur10} Curran~P.A. et al., 2010, MNRAS, arXiv:1007.5430


\bibitem[\protect\citeauthoryear{Fender et al.}{1999}]{fen99} Fender~R.P. et al., 1999, MNRAS, 304, 865

\bibitem[\protect\citeauthoryear{Fender et al.}{2004}]{fen04} Fender~R.P. et al., 2004, MNRAS, 355, 1105
\bibitem[\protect\citeauthoryear{Fender}{2006}]{fen06} Fender~R.P., 2006, in Lewin~W.H.G., van~der~Klis~M., eds., Compact Stellar X-ray Sources, Cambridge Univ Press, Cambridge, p.~381

\bibitem[\protect\citeauthoryear{Fender et al.}{2009}]{fen09} Fender~R.P. et al., 2009, MNRAS, 396, 1370

\bibitem[\protect\citeauthoryear{Hjellming \& Rupen}{1995}]{hje95} Hjellming~R.M., Rupen~M.P., 1995, Nat, 375, 464

\bibitem[\protect\citeauthoryear{Hjellming et al.}{1998}]{hje98} Hjellming~R.M. et al., 1998, BAAS, 30, 1405

\bibitem[\protect\citeauthoryear{Homan}{2010}]{hom10} Homan~J., 2010, Astronomer's Telegram, 2387

\bibitem[\protect\citeauthoryear{Kaaret et al.}{2003}]{kaa03} Kaaret~P. et al., 2003, ApJ, 582, 945

\bibitem[\protect\citeauthoryear{Markwardt et al.}{2009}]{mar09} Markwardt~C.B. et al., 2009, Astronomer's Telegram, 2258

\bibitem[\protect\citeauthoryear{Mirabel \& Rodr\'iguez}{1994}]{mir94} Mirabel~I.F., Rodr\'iguez~L.F., 1994, Nat, 371, 46
\bibitem[\protect\citeauthoryear{Mirabel \& Rodr\'iguez}{1999}]{mir99} Mirabel~I.F., Rodr\'iguez~L.F., 1999, ARA\&A, 37, 409

\bibitem[\protect\citeauthoryear{Miller-Jones et al.}{2004}]{mil04} Miller-Jones~J.C.A. et al., 2004, ApJ, 603, L21

\bibitem[\protect\citeauthoryear{Miller-Jones et al.}{2007}]{mil07} Miller-Jones~J.C.A. et al., 2007, MNRAS, 375, 1087

\bibitem[\protect\citeauthoryear{Miller-Jones et al.}{2008}]{mil08} Miller-Jones~J.C.A. et al., 2008, in Bandyopadhyay~R.M. et al., eds, A Population Explosion: The Nature and Evolution of X-ray Binaries in Diverse Environments, AIP Conf. Proceedings, 1010, 50

\bibitem[\protect\citeauthoryear{Nakahira et al.}{2010}]{nak10} Nakahira~S. et al., 2010, PASJ, arXiv:1007.0801

\bibitem[\protect\citeauthoryear{Shepherd et al.}{1994}]{she94} Shepherd~M.C. et al., 1994, BAAS, 26, 987

\bibitem[\protect\citeauthoryear{Shaposhnikov et al.}{2010}]{sha10} Shaposhnikov~N. et al., 2010, ApJ, arXiv:1008.0597

\bibitem[\protect\citeauthoryear{Szomoru}{2008}]{szo08} Szomoru~A., 2008, in Proceedings of the 9th EVN Symposium on The Role of VLBI in the Golden Age for Radio Astronomy and EVN Users Meeting, Proceedings of Science, PoS\,040

\bibitem[\protect\citeauthoryear{Tingay et al.}{1995}]{tin95} Tingay~S.J. et al., 1995, Nat, 374, 141


\end{thebibliography}
\end{document}